# Triple Helix indicators of knowledge-based innovation systems

## Introduction to the special issue

Loet Leydesdorff [1] & Martin Meyer [2]

**Abstract**

When two selection environments operate upon each other, mutual shaping in a co-evolution along a particular trajectory is one possible outcome. When three selection environments are involved, more complex dynamics can be expected as a result of interactions involving bi-lateral and tri-lateral relations. Three selection environments are specified in the Triple Helix model: (1) wealth generation (industry), (2) novelty production (academia), and (3) public control (government). Furthermore, this model somewhat reduces the complexity by using university-industry-government relations for the specification of the historical conditions of the non-linear dynamics. Whereas the historical analysis informs us about how institutions and institutional arrangements carry certain functions, the evolutionary analysis focuses on the functions of selection environments in terms of outputs. One can no longer expect a one-to-one correspondence between institutions and functions; a statistics is needed for the evaluation of how, for how long, and to what extent institutional arrangements enhance synergies among different selection environments. The empirical contributions to this Triple Helix issue point in the direction of "rich ecologies": the construction of careful balances between differentiation and integration among the three functions.

[1] Amsterdam School of Communications Research (ASCoR), University of Amsterdam, Kloveniersburgwal 48, 1012 CX Amsterdam, The Netherlands; loet@leydesdorff.net; http://www.leydesdorff.net .

[2] SPRU - Science and Technology Policy Research, University of Sussex, Brighton, United Kingdom.

**Introduction**

The Triple Helix (TH) model emerged from a workshop on *Evolutionary Economics and Chaos Theory: New Directions in Technology Studies* (Leydesdorff & Van den Besselaar, 1994) organized with the intention of crossing the boundaries between institutional analysis of the knowledge infrastructure, on the one hand (e.g., Etzkowitz, 1994), and evolutionary analysis of the knowledge base of an economy, on the other (David & Foray, 1994; Nelson, 1994). How can co-evolution between the layers of institutional arrangements and evolutionary functions be conceptualized in relation to the division of innovative labor among both institutions and functions? (Fritsch, 2004; Fritsch & Stephan, 2005)

While the knowledge-based system can itself be considered as an outcome of interaction among different social coordination mechanisms—markets, knowledge production, and (public or private) governance at interfaces—the Triple Helix model of university-industry-government relations provides us with a heuristic for studying these complex dynamics in relation to developments in the institutional networks among the carriers. The coupling to the layer of institutional networks, that is, the knowledge infrastructure of a knowledge-based system, reduces the complexity because the historical conditions limit the range of possible options. The observable events enable the analyst to specify the systems of reference in terms of initial conditions: which industries are empirically involved, and which bodies of knowledge, and what are the relevant levels of governance?

The evolutionary analysis focuses on the functions of selection environments in terms of outputs, whereas the historical analysis informs us about how institutions and institutional arrangements carry these functions (Andersen, 1994). In the call for papers for the first Triple Helix conference, Etzkowitz and Leydesdorff (1995) formulated this tension between the historical and evolutionary perspectives as follows:

> "Three sources of variation have been acknowledged in technology studies: (1) industrial sectors differ with respect to their relations to the technologies that are relevant for the developments in those sectors (e.g., Pavitt 1984); (2) different technologies induce different patterns of innovation and diffusion (e.g., Freeman & Perez 1988; Faulkner & Senker 1994); (3) systems of innovation (e.g., national systems of innovation) integrate and differentiate the various functions differently (Lundvall 1988; Nelson 1993). The variations, however, are both functional and institutional. The functional communications can sometimes be codified in new institutional settings; the institutional sectors (public, private and academic) that formerly operated at arm's length are increasingly working together, with a spiral pattern of linkages emerging at various stages of the innovation process." (Etzkowitz & Leydesdorff, 1995, p. 15).

Unlike institutions, functions are not observable without taking a reflexive turn, that is, without some specification of selection environments in terms of expectations. For example, when universities assume the functions of an entrepreneur or, at other times, of a regional innovation organizer—in addition to their traditional institutional missions of higher education and academic research—one needs to define what will be considered as 'entrepreneurship' or as an 'innovation organizer' before one can proceed to the measurement.

In a complex arrangement, functions (expectations) can no longer be expected to correspond in a one-to-one relation with institutions (observations), and therefore a set of statistics is needed for the analysis. The uncertainty between the layers operates upon the uncertainty in the institutional delineations and at the interfaces among the different functions (Table 1).

|  | *Sub-dynamics* | | |
|---|---|---|---|
| *Functions* | Wealth generation | Novelty production | Normative control |
| *Carriers* | Industry—University—Government | | |

**Table 1**:
A (neo-)institutional versus an evolutionary appreciation of the Triple Helix model

At the research level and given a project, one is able to reduce the complexity by 'black-boxing' one uncertainty or the other (e.g., by using a *ceteris paribus* clause and/or by specifying a focus of the research). The Triple Helix framework provides an opportunity to relate the various perspectives. It allows, for example, for studies of changes in institutional arrangements from a neo-institutionalist perspective, that is, in terms of networked relations among institutions (Powell & DiMaggio, 1991), yet also allows for the analysis from a neo-evolutionary perspective, that is, with a focus on the changes in functions (output) which one expects to be carried by the new arrangements.

**The Triple Helix as a heuristic**

From an evolutionary perspective, the institutions and the arrangements inform us about the retention mechanisms, that is, the fingerprints left behind by the complex dynamics. The TH model enriches the institutional analysis and makes the task of evolutionary reflection urgent. For example, when one focuses on university-industry relations, the addition of the dimension of government raises issues like the systemic evaluation of these relations.

In the analytical model, the different perspectives of government, industry, and academia can first be spanned along orthogonal axes, and the observables can then be appreciated as interaction effects among the functions (Figure 1). For example, since the second half of the 19$^{th}$ century, corporations have operated R&D laboratories, which contribute systematically to novelty production alongside academia. However, universities can also be considered as state apparatuses. Administrative innovations like patent legislation have reshaped the three environments by providing a nexus among them (Van den Belt & Rip, 1988).

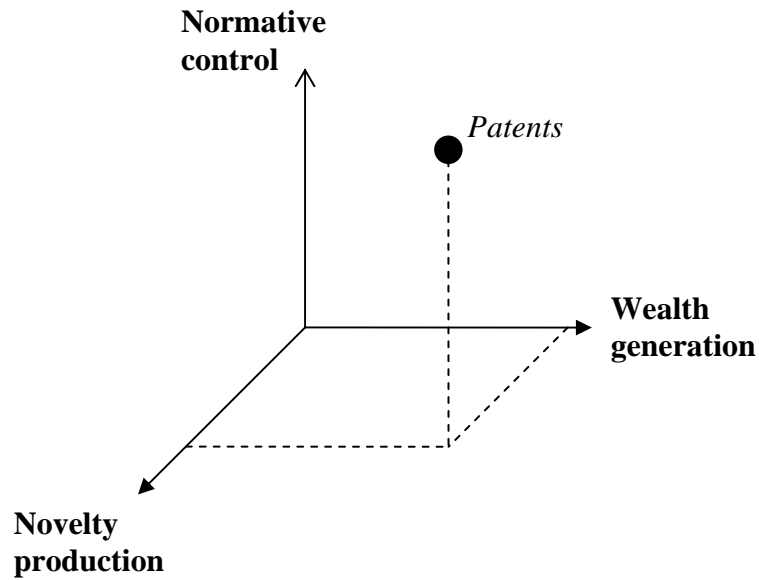

**Figure 1**:
An analytical scheme for studying the Triple Helix as a neo-evolutionary model

Figure 1 shows how observable units of analysis (for example, patents) can be appreciated from the three different perspectives. Patents are codified from the perspective of patent legislation—since patents need to be upheld in courts when litigated—but they are supposed to function both institutionally to secure revenues for the knowledge production process and also economically as investments in the value of intellectual property on relevant markets.

The three selection mechanisms operate upon each other's variation. In addition to the co-variation in the relations, the remaining variation in each dimension is important for assessing the relative significance of the respective co-variations. For example, universities produce more non-patentable knowledge than patentable inventions.

Industries use resources other than knowledge-intensive assets for production, and legislation other than patent legislation is relevant to the protection of intellectual property. The coordination mechanisms (markets, regulations, knowledge) co-vary in the observable events at each moment of time, and potentially co-evolve over time. The events (e.g., innovations) can be attributed as actions to the institutional carriers. At the functional level, however, the measurements update the expectations in terms of relative frequency distributions. Thus, the observations can be used as indicators with theoretically specified meanings in different selection mechanisms.

**Indicators and evolutionary dynamics**

This special issue of *Research Policy* is devoted to the study of Triple Helix indicators from the evolutionary perspective of studying the metrics of the new developments in terms that distinguish between expectations and observations. Systematic datasets enable analysts to position the conclusions of individual case studies: the case studies are considered as providing relevant hypotheses. In other words, this issue includes those contributions to the Fifth Triple Helix Conference (18-21 May 2005, Turin, Italy) that enable us to generate a critical distance from claims of new developments on the basis of single case studies or specific perspectives.

In principle, quantitative measurement can provide positive support to claims of a changing order, but the conclusions can be more refined. The analyst is able to specify the extent to which the relevant distributions changed, and whether new arrangements significantly affected the systems under study. Did these (sub)systems experience path-dependent transitions, or did the innovations under study only disturb an otherwise continuous development? The specification of the relevant systems of reference, and consequently the specification of criteria for the data collection with reference to evolutionary mechanisms, becomes crucial when studying these questions.

The issue is organized in several parts that correspond to different levels of aggregation and units of analysis. In the first part, the focus is on the systems level. Three studies assume different systems of reference: *Giovanni Dosi, Patrick Llerena, and Mauro Sylos Labini* discuss in their contribution—entitled "The relationships between science, technology, and their industrial exploitation"—the evaluation of indicators at the macro-level. The authors focus on what these indicators can teach us about the so-called "European Paradox", that is, the conjecture that EU countries play a leading role in terms of scientific output, but lag behind in their ability to convert this strength into wealth-generating innovations. By interpreting their data critically and from the perspective of evolutionary economics, the authors conclude that this European weakness is not confined to the exploitation of new knowledge, but is based on an increasing weakness of the knowledge production process itself. The authors argue for policy measures to strengthen frontier research both at universities and in the corporate sector.

The study by *Terry Shinn and Erwan Lamy* entitled "Paths of commercial knowledge: Forms and consequences of university-enterprise synergy in scientist-sponsored firms", demystifies another paradox: entrepreneurial scientists are not "blurring boundaries". On the contrary, with few exceptions, the highly specific contexts of research set boundaries that must be respected in order to be effective both at the research front and in the entrepreneurial arena. Thus, university-industry relations can be expected to have asymmetrical functions, which may be carried by different types of researchers. However, what these researchers seem to have in common is that they use their functions at interfaces as resource relations. The authors show that what seems at first sight to be a blurring of demarcations (Gibbons *et al*. 1994; Nowotny *et al*., 2001) can be considered as functional to the reproduction of the differentiation between the domains of academia and industry.

From a completely different (that is, economic) angle, *Gil Avnimelech and Morris Teubal* study the industry life-cycle of venture capital industries at the systems level. The authors argue that the process of emergence of this new industry and market has a dynamics of its own. This self-organizing dynamics facilitates specialization and division of knowledge both within the system of reference and in the co-evolving (high-tech) environments. The different (sub-)dynamics operate upon one another selectively, induce cycles, and accordingly strategic vectors on both sides of the interfaces. The more the variation is pre-structured, the better a selecting system can advance by developing its own dynamics.

**Knowledge spill-overs**

A second set of studies focuses on knowledge spill-overs in university-industry-government relations. *Pamela Mueller* explores "the knowledge filter" by considering university-industry relations and firm formation as drivers of economic growth at the regional level. Using a regression model, she concludes that "the proposed knowledge transmission channels—entrepreneurship and university-industry relations—increase the permeability of the knowledge filter, thus improving regional economic performance". In a paper entitled "The *ex ante* assessment of knowledge spill-overs: Government R&D policy, economic incentives and private firm behavior", *Maryann P. Felmann and Maryellen Kelley* present the results of an evaluation of the Advanced Technology Program (ATP) at the U.S. National Institute of Standards and Technology. The study concludes that the positive results of this governmental intervention "stand in stark contrast to the notion that government funding crowds out private investment". The authors note that statistically university linkages did not differ significantly between recipients of ATP funding and those who did not receive awards, since all the firms that applied for government funding had strong connections with universities already in place.

In summary, if university-industry relations and government interventions seem to prevail, the question arises of the conditions under which a knowledge-based economy emerges and how this can be indicated. In their paper entitled "Knowledge economy measurement: Methods, results and insights from the Malaysian knowledge content study", *Philip Shapira, Jan Youtie, K. Yogeesvaran, and Zakiah Jaafar* report on a comprehensive survey of Malaysian firms in both manufacturing and services. The authors find sector-specific knowledge content variables to be significant and recommend using the sector as the unit of analysis (Pavitt, 1984).

In their study entitled "Measuring the knowledge base of regional innovation systems in Germany in terms of a Triple Helix dynamics", *Loet Leydesdorff and Michael Fritsch* compare 438 German regions in terms of geographical distribution, size distribution, and the sectoral distribution of firms. Configurational information is proposed as a measure for the interaction effects among these three sources of variance. The authors conclude that the dynamics of knowledge-intensive services are different from those of high- and medium-tech manufacturing. Medium-tech manufacturing can be considered as the main driver of a knowledge-based configuration, while knowledge-intensive services tend to uncouple the local economy from its geographic basis. Policy implications from the perspective of promoting regional development are specified accordingly.

**The entrepreneurial university**

Oxfordshire's three universities and seven national/international research laboratories and their spin-offs are the focus of a study by *Helen Lawton Smith and Ka WaiHo* entitled "Measuring the performance of Oxford University, Oxford Brookes University and government laboratories' spin-off companies". This region is one of the most innovative of Europe, and Oxford University is one of the UK's most entrepreneurial universities. The authors argue that the survival rate of spin-off companies is high, but that it generally takes at least ten years before their rate of growth begins to accelerate. "Thus, spinning out companies is not a quick fix for government economic development strategies."

The embedding of innovative activities in a system of innovations is also a core conclusion of *Robert Tijssen*'s contribution entitled "University-industry interactions and university entrepreneurial science: Towards measurement models and indicators". Examining two indicators of connectivity—(1) public-private co-authored research articles, and (2) citations of university research articles in research articles with a corporate address—across a range of countries and sectors, the author concludes that these connectivity indicators are of minor significance compared to a university's country of location and the scale of its research activities in industrially relevant fields of science. One interesting finding is that patenting tends to be inversely related to the pursuit of a research partnership with industry. As the author argues, this result casts doubt on the appropriateness of using patent intensity as an indicator of university entrepreneurial performance.

In their study "Indicators and outcomes of Canadian university research: Proxies becoming Goals", *Cooper Langford, Jeremy Hall, Peter Josty, Stelvia Matos, and Astrid Jacobson* go one step further: on the basis of the results of a questionnaire, these authors warn against the use of proxies for statistical purposes without sufficient grounding in the reality of knowledge production and R&D. Most innovative activities take place under highly uncertain circumstances. According to these authors, institutional isomorphism in accordance with policy objectives may occur "if the wider knowledge base (in this case policy and academia) propagates the importance of proxies—i.e. if there is enough dogma, industry may follow suit". The authors argue for indicators which appreciate the time-lag in innovative activities and for the importance of on-going collaborations in graduate education as means to improve the innovative climate.

In another study of Canadian universities, *Réjean Landry, Nabil Amara, and Imad Rherrad* argue on the basis of a logistic regression of survey results "that the traditional and entrepreneurial visions of university research complement each other when looking at the resources mobilized by researchers to launch spin-offs". Out of all the investments that could be made, those related to the support of experienced researchers, the consolidation of social capital in networks, and a high degree of novelty of research knowledge have the largest marginal impacts on the likelihood of university spin-off creation.

In the rather different context of Italian universities, *Margherita Balconi and Andrea Laboranti* reach similar conclusions in their study entitled "University-industry interactions in applied research: The case of microelectronics": border-crossing collaborations tend to be driven by cognitive proximity and personal relationships. Strong relations are associated with high scientific performance by professors and are useful to firms for effective recruiting. The knowledge content can be considered as an endogenous driver of collaboration.

*Elefthérios Sapsalis, Bruno Van Pottelsberghe de la Potterie, and Ran Navon* generalize the points made above for the domain of academic patenting in their paper entitled "Academic *vs*. industry patenting: An in-depth analysis of what determines patent value". Their findings suggest that "Bayh-Dole Act-like regulations that occurred in the nineties in Europe have had a significant and beneficial impact". The boom in academic patents is associated with a value distribution similar to that of patents registered by the business sector. Self-citations in patents to scientific literature, that is, to the results of prior art, lead to patents of higher value. Patents thus provide a specific reflection of the knowledge-production process, while the latter has a dynamics of its own.

The dynamics of the patent production process in academia are also the subject of *Martin Meyer's* contribution, entitled "Are co-active researchers on top of their class? An exploratory comparison of inventor-authors with their non-inventing peers in nano-science and technology." The precise analysis of publication and patenting structures in this technology-driven field reveals internal differentiation within the scientific communication system: "While still over-represented among the highly cited authors, inventor-authors appear not to be among the most highly cited authors […]". However, according to this author, patenting does not have an adverse impact on publication or on the citation performance of researchers.

**The Triple Helix as an evolutionary model**

In the above summary of the contributions, we noted the tensions between integration and differentiation. Case studies tend to focus on institutional integration in terms of bilateral and trilateral relations. However, selection mechanisms in market environments are by nature very different from selection mechanisms within science or technology-like career-based reputations (Whitley, 1984). Different selection mechanisms can be expected to operate with other criteria, and this may stabilize or destabilize the developments along trajectories. Selections that operate upon selections can be expected to generate heavily skewed distributions. Under so much selection pressure, the previous integrations remain fragile and, therefore, one should proceed to normative conclusions with great care and taking all available information into account.

Unlike a Double Helix, which can be observed in DNA-molecules (Watson & Crick, 1953), a Triple Helix is only an analytical model which enables us to specify relevant categories for observation in terms of expectations. For analytical reasons, one cannot expect a Triple Helix to become stable and therefore observable (Li & Yorke, 1975; Leydesdorff, 2006a). The systems under study can be expected to remain in transition, and the observations have to be evaluated statistically (Etzkowitz & Leydesdorff, 1998; Leydesdorff & Etzkowitz, 1998). The observable arrangements inform us about the initial (historical) conditions or, in other words, the pathways selected by the evolving systems hitherto. However, the reflexive specification of the evolutionary dynamics in terms of selection environments may enable us to propose improvements in terms of the operating mechanisms. How can three sources of variance be expected to operate as selection environments for each other, and under what conditions can the interaction terms be used for innovations?

The biological model of Darwinian evolution theory assumed variation as the one sub-dynamic caused by random mutations, and selection as determined by nature. In evolutionary economics, one can distinguish among different selection environments. Variation can then be generated in the interactions among selection environments. However, different selection environments operate asymmetrically, that is, using different selection criteria: the observable variation can thus be considered as mutual information between the systems. The selection criteria (e.g., price/performance ratios, reputations) are system-specific. This leads to the question, "Which system is selecting in terms of what?"

In opposition to Nelson & Winter's (1982) evolutionary theory of economic change, Andersen (1994), for example, argued that not firms but coordination mechanisms (e.g., markets) are the evolving systems. Nelson (1994) proposed considering the relations between technologies and institutions in terms of co-evolution. In a co-evolutionary model, the two sub-dynamics are assumed to operate upon each other and thus the variation in the interaction is pre-structured (Dosi, 1982; McKelvey, 1996). A co-evolution may lead to "mutual shaping" (McLuhan, 1964) and "lock-in" (Arthur, 1988, 1989) if the co-variation between the two sub-dynamics at each moment of time is reinforced over time (e.g., by network externalities). When this reinforcement is considered as containing a third sub-dynamic—that is, no longer considered as a linear progression along the time axis—a Triple Helix model is generated analytically.

For example, university-industry arrangements in a small nation like the Netherlands were based in the 1950s on informal relations at the level of scholarly and engineering communities between a few large corporations (Philips, Unilever, Royal Dutch/Shell) and the (technical) universities (Leydesdorff *et al.*, 1980). The state supported these arrangements, but did not pursue an active S&T policy at that time. S&T policies emerged in the 1960s and 1970s alongside industrial policies and policies for higher education. From the very beginning, these S&T policies were not strictly national (Yamauchi, 1986; Freeman, 1987; Lundvall, 1988). For example, they were induced at the level of the OECD, and they were increasingly shaped with reference to an emerging European level (Vavakova, 2000). Thus, national government could no longer be considered as a constant in the institutional arrangements, but multi-level *governance* became itself increasingly a source of variation (Kooiman, 1993; Leydesdorff, 2005).

A system with three sub-dynamics is complex and, in principle, may contain all kinds of chaotic behavior like bifurcations, crises, strange attractors, etc. When the three sub-dynamics are not synchronized *ex ante*, they can be considered as analytically independent sources of variance. However, one can expect co-evolutions to develop continuously between each two of them, that is, at interfaces (Kline & Rosenberg, 1986). Co-evolutions develop along trajectories with the possibility of "lock-in." A system with three sub-dynamics selects upon these trajectories like a regime, given the historical configuration. Thus, another cycle is added (Schumpeter, 1939; Abernathy & Clark, 1985; Freeman & Perez, 1988).

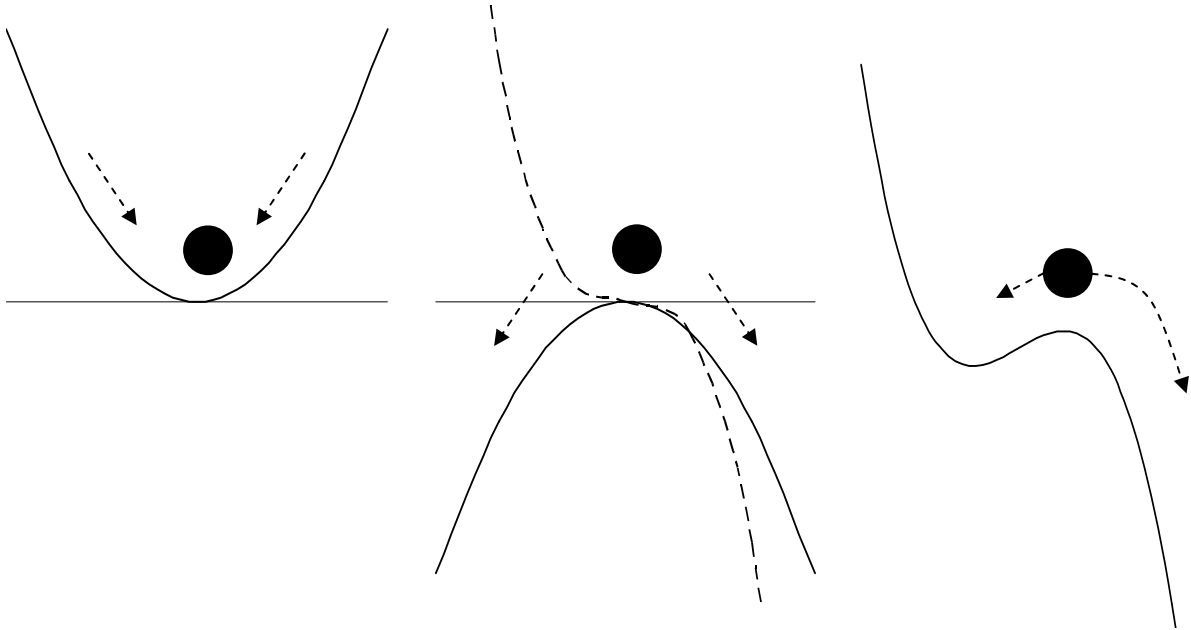

**Figure 2a, b, and c.** The possibility of destabilization, meta-stabilization, and globalization by adding a third selection environment to a hitherto stabilized trajectory.

While co-evolutions between selection environments can stabilize technological trajectories in processes of mutual shaping, the additional degree of freedom in a system with three sub-dynamics may lead to de-stabilization, meta-stabilization, and globalization (Figure 2). For example, when decreasing marginal return are replaced with increasing marginal returns—as some authors have suggested for the case of information technologies and services (Arthur, 1988; Barras, 1990)—the valley provided for the trajectory at the minimum of the hyperbola (Figure 2a; Sahal, 1985) can become a hilltop in the model given the change in the sign of the feedback (Figure 2b). At the saddle-point of the three-dimensional curve (the dashed line in Figure 2b), a system can be expected to bifurcate. When the three selection mechanisms operate with different parameters (Figure 2c), the resulting system will tend to be locked-in to sub-optima. However, if the hill is

climbed—using a trajectory—the system can also reach another basin of attraction (Leydesdorff, 2006b).

From this abstract perspective of complex-systems theory, a co-evolution along a trajectory reduces complexity and thus enables the system to move from one basin of attraction to another. For example, a dominant design in aircraft development like the DC3 (1936) generated first the regime of civil aviation, while the development of wide-body aircrafts with jet engines (since the introduction of the Boeing 707 in 1957) generated the regime of transatlantic mass-transportation and tourism using charter planes in the decades thereafter (Frenken & Leydesdorff, 2000; Sahal, 1985). Historical conditions force us to reduce the analytically possible complexity along a contingent trajectory that can serve competitors as a pathway during hill-climbing. When a next plateau is reached, the old regime is left behind and the relevant selection environments are redefined (Teubal, 1979). For example, when the VCR was replaced with the DVD, the previous lock-in of VHS versus Betamax (Arthur, 1988) rapidly became irrelevant.

**The evolutionary drive in Triple Helix transitions**

In an economic model of this complex system, the attractors of development are markets (or non-market selection environments), because these diffusion dynamics drive the systems with expectations of profit. However, diffusion dynamics may destabilize a hitherto coevolving system—for example, of production and distribution (Callon *et al*., 2002)—when the diffusion parameter becomes relatively large (Rashevsky, 1940; Turing, 1952). A saddle point (the dashed line in Figure 2b) is then generated in the phase space and thus a bifurcation can be expected.

For example, in regional innovation systems like Italian districts, a co-evolution between the knowledge-production function and local markets can first carry an innovation along its trajectory. As the innovation matures and volume is generated on the market, an additional diffusion dynamics may become relevant. For example, a multinational corporation may buy the local firm that produces the innovated products, and then relocate production facilities for geographical reasons. The diffusion dynamics thus dissolves the previous co-evolution along a trajectory in a next stage. The innovation that was previously stabilized in a specific region, can thus be globalized at the level of the market.

The innovative district remains under this threat of de-industrialization because the innovated system can be expected to contain a dynamics different from the innovating one (Beccatini, 2003). Consequently, new innovations would continuously have to be generated. A specific combination of local institutions and the knowledge-production function (e.g., in Silicon Valley) may be sufficiently complex to counter this de-industrialization threat with further innovations (Nowak & Grantham, 2000; Etzkowitz & Klofsten, 2005; Cooke & Leydesdorff, 2006). However, this need of continuous innovations drives the system(s) into becoming knowledge-based. The knowledge-based systems transform the institutional conditions increasingly into a knowledge infrastructure or, in other words, a Triple Helix network of university-industry-government relations.

**Normative implications**

Along the lines of the "Mode 2" thesis (Gibbons *et al*., 1994), the Triple Helix model has sometimes been understood as a plea for blurring the boundaries between universities, industry, and government (e.g., Raman, 2005). In our opinion, this normative use of the Triple Helix model can be appreciated with hindsight as a specific reaction to the historical configuration that emerged in the early 1990s when the effects of the Bayh-Dole Act made it necessary for European countries to consider new legislation about university patenting (Sapsalis *et al*., 2006). In the meantime, however, patenting by universities and university staff has reached a stable level in the U.S.A. (Figure 3; cf. Mowery & Ziedonis, 2002).

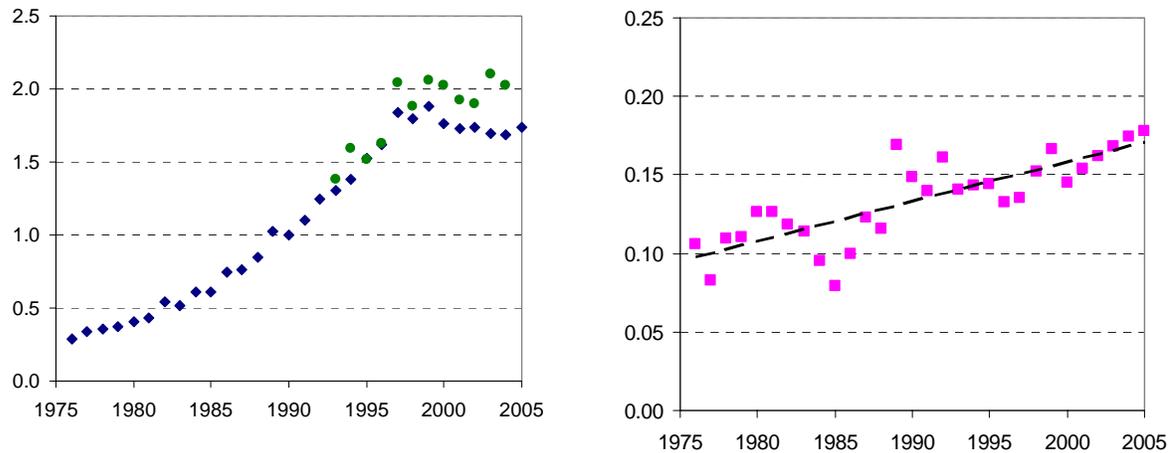

**Figure 3**: Percentage of U.S. patents with the words "University" (♦) and "Institute of Technology" (■) in the address field of the assignees. Source: USPTO database. Percentage of U.S. patents issued (●) on the basis of survey results (AUTM, 2005, at p. 16).

As an evolutionary model, the Triple Helix abstracts from the institutional premises of a specific period. By considering the three selection environments as functions in the research design, the knowledge-based innovation systems can be studied as rich ecologies. Rich ecologies are based on careful balances between differentiation and integration. More than normatively inspired and *ex ante* calls for integration, empirical studies based on evolutionary models enable agents to distinguish integration from differentiation in different dimensions, and thus to contribute reflexively to the construction of competitive advantages in the knowledge-based economy by specifying new functions and institutional needs.

**Acknowledgement**

We are grateful to Terry Shinn for his comments on a previous draft and would like to thank Jenny Newton for her editorial assistance.